\def\year{2022}\relax
\documentclass[letterpaper]{article} 
\usepackage{aaai22}  
\usepackage{times}  
\usepackage{helvet}  
\usepackage{courier}  
\usepackage[hyphens]{url}  
\usepackage{graphicx} 
\usepackage{natbib}  
\usepackage{caption} 
\DeclareCaptionStyle{ruled}{labelfont=normalfont,labelsep=colon,strut=off} 
\usepackage{mathptmx}
\usepackage{comment}

\usepackage{hyphenat}
\usepackage{amsmath}
\usepackage{color}

\usepackage{xspace}
\usepackage{subcaption}
\usepackage{footmisc}

\newcommand{\abbrevStyle}[1]{#1}

\newcommand{\ie}{\abbrevStyle{i.e.}\xspace}
\newcommand{\eg}{\abbrevStyle{e.g.}\xspace}
\newcommand{\cf}{\abbrevStyle{cf.}\xspace}

\newcommand{\vs}{\abbrevStyle{vs.}\xspace}

\newcommand{\Eqnref}[1]{Eq.~\ref{#1}}

\newcommand{\Figref}[1]{Fig.~\ref{#1}}

\newcommand{\xhdr}[1]{\vspace{1.7mm}\noindent{{\bf #1.}}}

\newcommand{\textcite}[1]{\citeauthor{#1} \shortcite{#1}}

\newcommand{\cpt}[1]{\textsc{\MakeLowercase{#1}}}

\newcommand{\hide}[1]{}

\usepackage{xcolor}
\newcommand{\answerYes}[1]{\textcolor{blue}{#1}} 
 
\newcommand{\answerNA}[1]{\textcolor{gray}{#1}} 
 
\title{Curious Rhythms: Temporal Regularities of Wikipedia Consumption}

\author {
    Tiziano Piccardi\textsuperscript{\rm 1}\footnote{Work done while at EPFL.},
    Martin Gerlach\textsuperscript{\rm 2},
    Robert West\textsuperscript{\rm 3}\footnote{Robert West is a Wikimedia Foundation Research Fellow.}
}
\affiliations {
     \textsuperscript{\rm 1}Stanford University\\
    \textsuperscript{\rm 2}Wikimedia Foundation\\
    \textsuperscript{\rm 3}EPFL\\
    piccardi@stanford.edu, mgerlach@wikimedia.org, robert.west@epfl.ch
}

\begin{document}

\maketitle

\begin{abstract}

Wikipedia, in its role as the world's largest encyclopedia, serves a broad range of information needs.
Although previous  studies have noted that Wikipedia users' information needs vary throughout the day, there is to date no large-scale, quantitative study of the underlying dynamics.
The present paper fills this gap by
investigating temporal regularities in daily consumption patterns in a large-scale analysis of billions of timezone-corrected page requests mined from English Wikipedia's server logs, with the goal of investigating how context and time relate to the kind of information consumed. First, we show that even after removing the global pattern of day--night alternation, the consumption habits of individual articles maintain strong diurnal regularities. Then, we characterize the prototypical shapes of consumption patterns, finding a particularly strong distinction between articles preferred during the evening/night and articles preferred during working hours. 
Finally, we investigate topical and contextual correlates of Wikipedia articles' access rhythms, finding that article topic, reader country, and access device (mobile \vs\ desktop) are all important predictors of daily attention patterns. 
These findings shed new light on how humans seek information on the Web by focusing on Wikipedia as one of the largest open platforms for knowledge and learning, emphasizing Wikipedia's role as a rich knowledge base that fulfills information needs spread throughout the day, with implications for understanding information seeking across the globe and for designing appropriate information systems.
\end{abstract}

\section{Introduction}

Human life is driven by strong temporal regularities at multiple scales, with  events and activities recurring in daily, weekly, monthly, yearly, or even longer periods \cite{kreitzman2011rhythms}.
The cyclical nature of human life is also reflected in human behavior on digital platforms.
Understanding and modeling temporal regularities in digital\hyp platform usage is important from a practical as well as a scientific perspective:
on the practical side, understanding user behaviors and needs is critical for building effective, user-friendly platforms;
on the scientific side, since online behavior reflects user needs, studying temporal regularities of platform usage can shed light on the structure of human life itself and is thus consequential for sociology, psychology, anthropology, medicine, and other disciplines.
For instance, in information science, time has long been recognized as a crucial contextual factor that drives human information seeking \cite{savolainen2006time}.
The study of traces recorded on digital platforms has yielded novel insights about circadian rhythms \cite{aledavood2022quantifying,pinter2022awakening,althoff2017harnessing,murnane2015social,chalmers2011rhythms,kumar2020habitual,leypunskiy2018geographically} and periodic fluctuations in alertness \cite{abdullah2016cognitive}, mood \cite{dodds2011temporal,golder2011diurnal}, focus \cite{mark2016sleep}, musical taste \cite{park2019global,heggli2021diurnal}, purchase habits \cite{gullo2019does}, and ad engagement \cite{saha2021advertiming}.

With regard to both aspects---practical and scientific---Wikipedia constitutes a particularly important online platform to study:
On the practical side, Wikipedia is one of the most frequently visited sites on the Web, such that a better understanding of user behavior can potentially lead to site improvements with consequences for billions of users.
On the scientific side, Wikipedia is not just yet another website; it is the world's largest encyclopedia, where each page is about a precise concept.
Temporal regularities of Wikipedia usage thus have the potential to reveal regularities of human necessities, telling us what humans care about, and what they are curious about, at what times.
Some temporal regularities of Wikipedia usage are already known:
\eg, Wikipedia's overall usage frequency, as well as the length of sessions, varies by time of day \cite{piccardi2023large}, as do users' reasons for reading Wikipedia \cite{singer_why_2017,lemmerich_why_2019}.
These studies have, however, glanced at temporal regularities merely superficially and in passing while focusing primarily on other aspects of Wikipedia usage.
The most related to our work is a study that showed that Wikipedia \textit{editing} follows daily and weekly regularities \cite{yasseri_dynamics_2012}, but the study was limited to editing behavior.
One reason why previous studies have not been able to analyze reading, rather than editing, behavior to date is that, as opposed to edit logs (where editors' geo-locations can be approximated via logged IP addresses), geo-located reading logs are not publicly available.
Rather, Wikipedia's public hourly pageview logs\footnote{\url{https://dumps.wikimedia.org/other/pageviews/readme.html}} report only Coordinated Universal Time (UTC) without specifying the reader's local time.
Especially in large language editions such as English, which are accessed from many countries in different timezones, this constitutes a major limitation.

We overcome this limitation by working with a non-public dataset of English Wikipedia's full webrequest logs, enriched with timezone information inferred from user IP addresses, which allows us to timezone-correct all timestamps and thus to faithfully study, for the first time, temporal regularities of Wikipedia reading at hourly granularity.
We characterize how information consumption on the platform varies by time of day and how time interacts with other contextual properties, such as article topics and reader country.
Adding to previous studies on the consumption and popularity of Wikipedia content, we provide insights into the daily temporal rhythms that drive Wikipedia usage. 
Specifically, we address the following research questions:
\begin{description}
    \item[RQ1] \textbf{{Strength of rhythms:}}  How strongly is Wikipedia consumption driven by periodic rhythms?
    \item[RQ2] \textbf{{Shapes of rhythms:}} What are the typical shapes of Wikipedia consumption rhythms?
    \item[RQ3] \textbf{{Correlates of rhythms:}} How do topical and contextual factors determine Wikipedia consumption rhythms?
\end{description}

Regarding RQ1, we find that fluctuations in Wikipedia's total access volume are largely explained by a diurnal baseline rhythm corresponding to the human circadian wake--sleep cycle.
Crucially, however, individual articles deviate systematically from the baseline rhythm, with deviations themselves following periodic diurnal patterns.

Regarding RQ2, principal component analysis reveals that individual articles' access rhythms are heavily driven by a small number of prototypical temporal signatures, but that article\hyp specific rhythms do not cluster into distinct groups, but vary smoothly along a continuum.

Regarding RQ3, regression analysis shows that temporal regularities in article access volume vary systematically by article topic, access method (mobile \vs\ desktop), and reader country. The latter is the strongest determinant of access rhythm, whereby different countries' interests in English Wikipedia fluctuate in idiosyncratic periodic patterns over the course of the day.
In terms of topic, to a first approximation, \cpt{STEM} and \cpt{History \& Society} articles tend to be more popular early in the day, and \cpt{Media} articles later in the day, with \cpt{Culture} articles being less concentrated in time.

Taken together, these findings further our understanding of how human information needs vary in space and time, and can serve as a stepping stone toward the informed design of improved information systems, on Wikipedia and beyond.

\section{Related Work}
\label{sec:related_work}

\xhdr{Temporal rhythms in digital traces}
Temporal patterns using digital traces have been explored in a large variety of topics. Consumption logs can bring invaluable insight into our understanding of how time impacts human activities. For example, mobile phone traces can help characterize daily rhythms \cite{aledavood2022quantifying} and expose insights about our circadian rhythm \cite{pinter2022awakening}, which may be difficult to study otherwise.
Analyzing the temporal dynamics of digital traces can also uncover biological and cognitive rhythms during the day, such as sense of alertness \cite{abdullah2016cognitive}, chronotypes \cite{althoff2017harnessing}, wellness \cite{zhou2023circadian}, and focus \cite{mark2016sleep}.
Individual temporal rhythms can also serve as predictors of medical conditions that need attention \cite{doryab2019modeling}.
Social media data, like Twitter \cite{murnane2015social,chalmers2011rhythms,grinberg2013extracting}, and messaging apps, like WhatsApp \cite{kumar2020habitual}, can be used as large-scale sensors revealing insights about sleeping patterns \cite{leypunskiy2018geographically}, and happiness fluctuations \cite{dodds2011temporal}, with variations across cultures \cite{golder2011diurnal}.
Time of day also impacts our choices in terms of musical taste \cite{park2019global,heggli2021diurnal}, purchase habits \cite{gullo2019does}, and ad engagement \cite{saha2021advertiming}.

Interactions with Wikipedia, too, are affected by time. Previous studies describe how consumption patterns reveal information about seasonal fluctuations in mood \cite{dzogang2016seasonal} at the population scale and how editors' temporal patterns exhibit culture\hyp specific differences \cite{yasseri2012circadian}. It has also been noted that modeling Wikipedia temporal readers' needs has implications for technical improvements and could be exploited to develop scalable infrastructure adapted to reading patterns \cite{reinoso2012characterization}. 

\xhdr{Wikipedia reader behavior} 
Given Wikipedia's central role in the Web ecosystem, there is increasing attention to characterizing reader behavior. Previous work investigated the reasons that lead readers to consume Wikipedia content \cite{singer_why_2017}, finding differences by time and country \cite{lemmerich_why_2019}. 
Complementary studies focus on consumption dynamics \cite{ratkiewicz_characterizing_2010,ribeiro2020sudden,ReaderPreferences,DwellingTime} and interaction with various article elements, such as citations \cite{maggio2020meta,piccardi2020quantifying}, external links \cite{piccardi2021gateway}, and images \cite{rama2021large}.
Significant attention is also dedicated to navigating from page to page to understand the mechanism that allows users to move across content in a natural setup \cite{piccardi2023large,wikirabbithole,trattner_exploring_2012,helic_models_2013,singer_detecting_2014,StructureArticlesNavigation,cornelius_inspiration_2018,dimitrov2019different} or in lab-based settings, such as wiki games \cite{Wikispeedia,HumanWayfinding,Helic,West_Leskovec_2012,LastClick,PredictingNavigationSuccess}.
Our work complements these findings by focusing on the shape of temporal regularities, which have not been analyzed in detail before.

\xhdr{Information needs and curiosity} 
Our work relates broadly to more theoretical work on information needs. The formulation of information needs was initially popularised by \citeauthor{wilson1981user} \cite{wilson1981user} and the resulting information models refined over the years \cite{wilson1997information,wilson1999models}. 
Information needs have also been linked to concepts in biology by comparing the need for information to the need for food \cite{machlup1983study}. This idea is formalized in information foraging theory \cite{pirolli1999information}, which has developed behavioral models describing humans in the information space as predators relying on mechanisms such as information scent \cite{chi2001using} to find what they need \cite{suh2010want,mangel2013invasion,rotman2011slacktivism}.
More recent work \cite{Lydon-Staley2021,zhou2023architectural} investigates the mechanism of online consumption by focusing on the role of curiosity, using Wikipedia as the reference platform and finding substantial differences in how humans explore information networks.

\section{Data}
\label{sec:data}
Our study relies on the access logs of English Wikipedia, collected over four weeks (1--28 March 2021) on the servers of the Wikimedia Foundation. These server logs, stored for a limited time, describe the requests received by the server when readers access the site, capturing the title of the requested article, the time the request was made, the user's IP address and geo\hyp location (approximately inferred from the IP address), and more.

In this study, we focus only on the traffic generated by human readers by discarding all bots. In order to ensure anonymity, we prepare the data as follows. First, we consider only requests for articles (namespace~0), ignoring requests for pages in other namespaces (\eg, talk pages), and we consider only requests originating from external websites, ignoring requests made from other Wikipedia pages. 
The restriction to incoming traffic from external websites better captures the (exogenous) information needs that cause users to visit Wikipedia, rather than (endogenous) information needs that are caused by visiting Wikipedia.
Additionally, we refine the logs by removing sequential loads of the same page from the same client because they could be an artifact of the browser \cite{piccardi2023large}, not representing a real intention to visit Wikipedia.

We anonymize the data by removing all sensitive information such as IP addresses, user-agent strings, fine\hyp grained geo\hyp coordinates, and all requests from logged-in users (3\% of the pageloads). Finally---and crucially for our purposes---we align all requests by converting timestamps to the user's local time using timezone information available in the logs. 

After these steps, we retain 3.45B pageloads associated with 6.3M articles. 
We represent the number of pageloads of article $a$ in hour $h$ of the week (averaged over the four weeks) by $n_a(h)$, \ie, each article is represented by a 168-dimensional time series, with one entry for each of the $168 = 24 \times 7$ hours per week.

\xhdr{Article properties} ORES \cite{ORES}, Wikipedia's official article scoring system, offers a classifier for labeling articles with topics. The labels are organized in a two-level taxonomy manually derived from WikiProjects\footnote{\url{https://en.wikipedia.org/wiki/Wikipedia:WikiProject}} (groups of editors who self-organize to work on specific topical areas). 
We applied the classifier to all articles in our dataset and, for each article, obtained a probability for each of 38 topics, grouped into five top-level topics: \cpt{STEM}, \cpt{Culture}, \cpt{History \& Society}, \cpt{Media} (originally included in \cpt{Culture}, but promoted to a top-level topic for our study), \cpt{Geographical}.
Note that each article is independently scored for each topic, so topic probabilities do not sum to 1 for a given article.

\xhdr{Access properties} For each pageload, we retain two contextual properties of the request: access method and user country. The access method indicator specifies if the user loaded the article from a desktop or a mobile device, such as a smartphone or tablet. The user country
is estimated by geo-locating the IP address associated with the request.

\begin{figure}[t]
    \centering
    \includegraphics[width=0.99\linewidth]{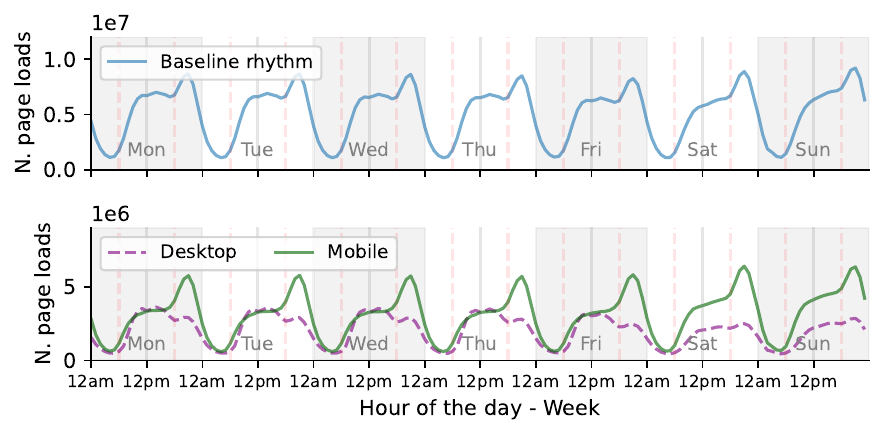}
    \caption{\textit{Top:} Total number of pageloads from external origin by hour of the week, averaged over four weeks. \textit{Bottom:} Idem, stratified by desktop and mobile.}
    \label{fig:weekly_pattern_landing}
\end{figure}

\xhdr{Baseline rhythm}
\Figref{fig:weekly_pattern_landing} (top) shows the total number of pageloads per hour of the week (averaged over the four weeks considered). 
The consumption rhythm follows a diurnal pattern, with the lowest access volume between 4:00 and 5:00 from Monday to Saturday and between 5:00 and 6:00 on Sunday, and with a sharp increase in traffic from 18:00 every day of the week, consistently reaching its peak at 21:00.
When broken down by access method (\Figref{fig:weekly_pattern_landing}, bottom), different device types are associated with different patterns, with mobile devices driving the increase in evening traffic. Similarly, access from desktop devices, likely more dependent on working rhythms, shows a small reduction around 12:00 and 18:00 and reduced activity on weekend mornings.

Denoting the number of pageloads in hour $h$ by $N(h) = \sum_a n_a(h)$ and normalizing $N(h)$ to be a distribution over the 168 hours of a week, we obtain what we term Wikipedia's \textit{baseline rhythm} $\Pr(h) := N(h)/\sum_{h'=0}^{167}N(h')$.

\xhdr{Divergence from the baseline rhythm}
In addition to Wikipedia's overall access volume (\Figref{fig:weekly_pattern_landing}),
we are interested in studying temporal access volume regularities for individual articles $a$, by analyzing article\hyp specific distributions over hours, $\Pr(h|a) := n_a(h)/\sum_{h'=0}^{167}n_a(h')$.
Individual articles' rhythms $\Pr(h|a)$ are heavily driven by the overall baseline rhythm $\Pr(h)$ dictated by human wake--sleep patterns.
Therefore, to study article\hyp specific rhythms in isolation from the baseline rhythm, we remove the latter by computing the hourly \textit{divergence} $D_a(h)$ of $a$'s rhythm $\Pr(h|a)$ from the baseline rhythm $\Pr(h)$ via pointwise division:
\begin{equation}
\label{eqn:baseline_removal}
D_a(h)
=\frac{\Pr(h|a)}{\Pr(h)}
=\frac{n_a(h)/\sum_{h'=0}^{167}n_a(h')}{N(h)/\sum_{h'=0}^{167}N(h')}.
\end{equation}
A divergence greater [less] than 1 indicates that in hour $h$, article $a$ receives more [less] attention than expected from the baseline rhythm, whereas a divergence of 1 implies that a global weekly rhythm fully determines $a$'s consumption.

Given the skewed distribution of article popularity, to avoid data sparsity issues, we keep only articles with at least 1,000 pageloads over the four weeks considered, or 35 pageloads per day on average. 
After applying this filter, we retain 439K articles,
accounting for 83.8\% (2.89B pageloads) of English Wikipedia's traffic.

\section{RQ1: Strength of Rhythms}
\label{sec:RQ1}
We start by quantifying the strength of temporal regularities in Wikipedia consumption patterns.
Taking a signal processing approach, we decompose a time series $x(h)$ of interest (\eg, $x(h)=\Pr(h)$ or $D_a(h)$) into its frequency components via the Fourier transform, which represents $x(h)$ as a weighted sum of sinusoids of all possible frequencies.
We denote the weight (Fourier coefficient) of frequency $f \in \{0, \dots, 167$\} by $\hat x(f)$.
We then measure the contribution of each frequency $f$ to time series $x(h)$ (or, equivalently, to $\hat x(f)$) via the so-called \textit{energy spectral density} $E(f)$, which captures the fraction of $x(h)$'s total variance (or energy) explained by each frequency~$f$:
\begin{equation}
E(f) = \frac{|\hat{x}(f)|^2}{\sum_{f'=0}^{167}|\hat{x}(f')|^2}.
\end{equation}

%

\begin{figure}[t]
    \centering
    \includegraphics[width=0.95\linewidth]{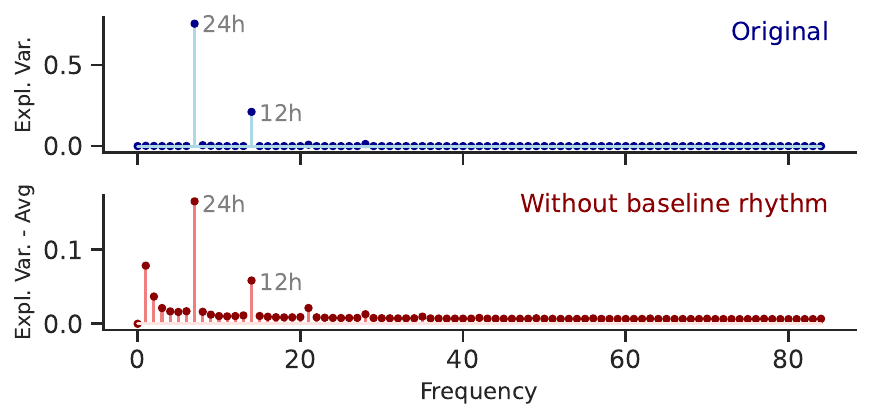}
    \caption{\textit{Top/blue:} Contribution of each frequency to baseline rhythm $\Pr(h)$ of Wikipedia access volume  (measured as the fraction of total variance explained).
    \textit{Bottom/red:} Contribution of each frequency to article\hyp specific divergence $D_a(h)$ from baseline rhythm (\Eqnref{eqn:baseline_removal}) (computed per article $a$, then averaged over articles).
    }
    \label{fig:pattern_energy}
\end{figure}

\xhdr{Baseline rhythm} 
Using $x(h)=\Pr(h)$ lets us study temporal regularities in the weekly baseline rhythm.
\Figref{fig:pattern_energy} (top) plots the variance $E(f)$ explained by each frequency component $f$ of the baseline rhythm $\Pr(h)$ (mean\hyp centered across $h$ in order to discard the constant offset term corresponding to $f=0$).
As previously anticipated by \Figref{fig:weekly_pattern_landing}, the dominant frequency corresponds to a 24-hour cycle ($f=7$). 
The daily and half-daily cycles ($f=7$ and 14; wavelengths 24 and 12 hours)  together explain 96.2\% of the variance (74.8\% and 21.4\%, respectively).
When split by access method, the pattern on mobile is more predictable, with daily and half-daily cycles explaining 94\% of the total variance for mobile, \vs\ 87\% for desktop.

\xhdr{Article-specific rhythms} To measure the strength of article\hyp specific daily rhythms that do not depend on the overall baseline rhythm $\Pr(h)$, we next repeat the above analysis, but now with $x(h) = D_a(h)$ (again mean\hyp centered across $h$).
We obtain the energy spectral density $E(f)$ for each article $a$ and average it over $a$ in order to measure the average strength of frequency $f$ in divergences $D_a(h)$ from the baseline rhythm, plotted in
\Figref{fig:pattern_energy} (bottom).
We see that, even after removing the overall baseline rhythm, still 22.3\% of the variance is explained by a combination of the cycles of wavelengths of 24 (16.5\%) and 12 (5.8\%) hours.
This means that the individual articles' access volume rhythms are not just driven by the change in the overall number of pageloads throughout the day/week, but are also strongly determined by article\hyp specific factors.

\hfill

To summarize, investigating RQ1, we found that overall Wikipedia consumption follows a strong baseline day--night rhythm, but also that individual articles considerably deviate from this baseline rhythm.
The deviations themselves also largely follow diurnal patterns.

\begin{figure}[t]
    \begin{minipage}[t]{.49\columnwidth}
        \centering
        \includegraphics[width=3.9cm]{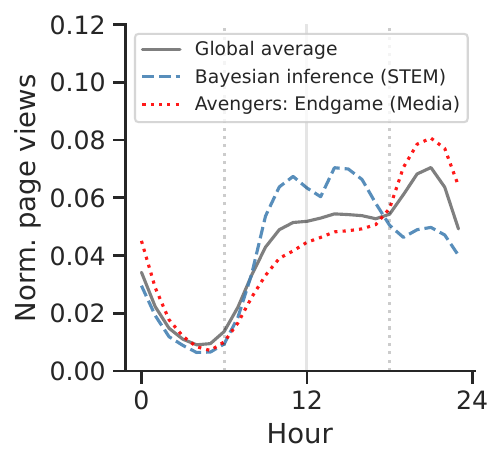}
        \subcaption{Daily pattern}
        \label{fig:STEMvsMovie_series}
    \end{minipage}
    \begin{minipage}[t]{.49\columnwidth}
        \centering
        \includegraphics[width=3.9cm]{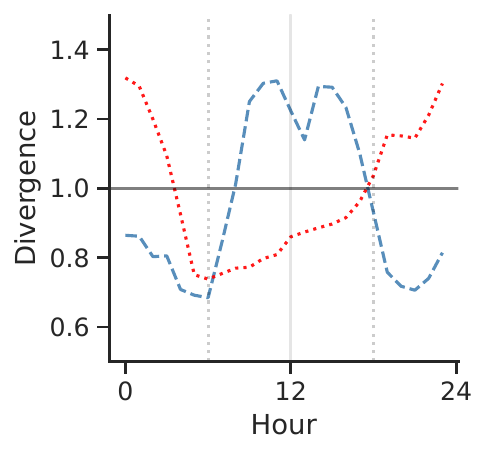}
        \subcaption{Divergence}
        \label{fig:STEMvsMovie_ratio}
    \end{minipage}
    \hfill
    
\caption{
Daily access volume of two articles with different topics: \cpt{STEM} (dashed blue) and \cpt{Media} (dotted red). (a) Normalized time series $\Pr(h|a)$. (b) Divergence $D_a(h)$ of $\Pr(h|a)$ from the overall baseline rhythm $\Pr(h)$ (\cf\ \Eqnref{eqn:baseline_removal}).
}
\label{fig:STEMvsMovie_example}
\end{figure}

\begin{figure}[t]
    \centering
    \includegraphics[width=0.95\linewidth]{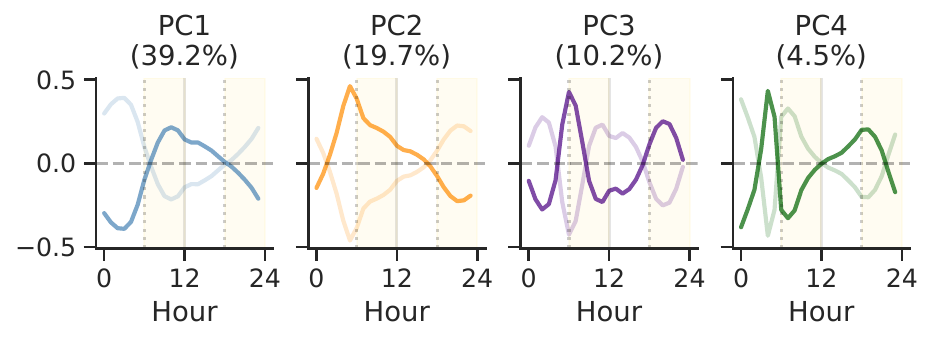}
    \caption{Four principal components of the daily access volume time series, capturing 73.6\% of total variance.}
    \label{fig:principal_components}
\end{figure}

\begin{figure*}[t]
\hfill
    \begin{minipage}[t]{.72\textwidth}
        \centering
        \includegraphics[height=3.9cm]{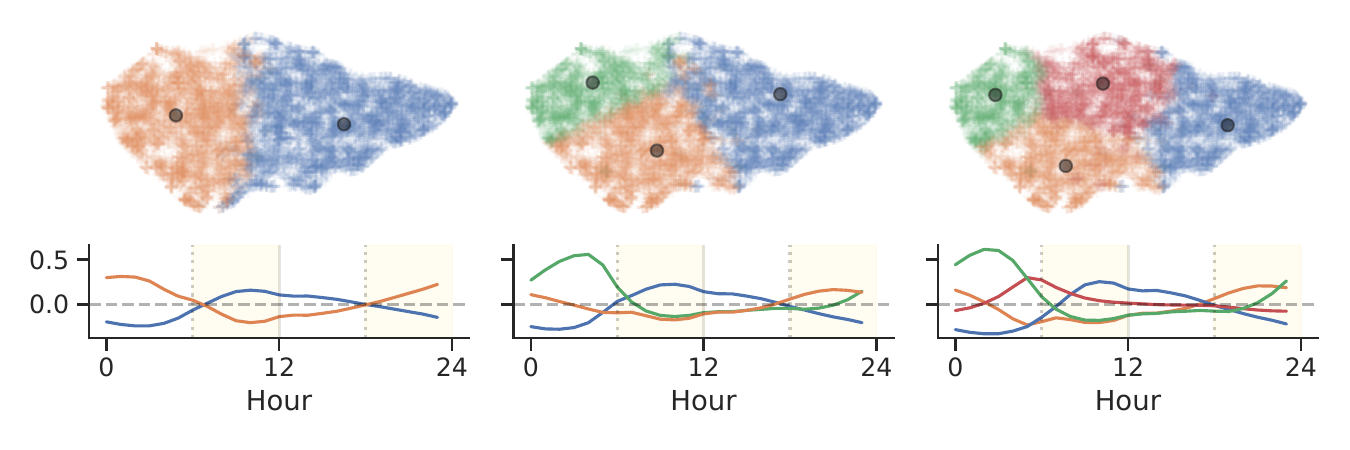}
        \subcaption{$k$-means clusters}
        \label{fig:clusters}
    \end{minipage}
    \begin{minipage}[t]{.24\textwidth}
        \centering
        \includegraphics[height=3.9cm]{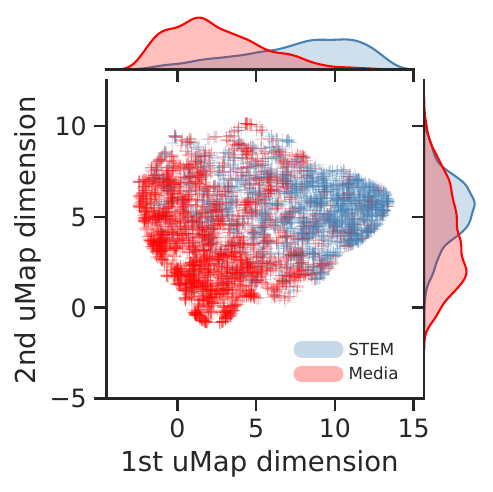}

        \subcaption{STEM vs. Media}
        \label{fig:STEMvsMedia}
    \end{minipage}
    \hfill
    
\caption{(a) \textit{Top:} UMAP projection of access volume time series for 10K random articles, with colors representing clusters obtained with $k$-means for different values of $k=2, 3, 4$. \textit{Bottom:} Centroids reconstructed from their first four principal components. (b) A subset of the same 10K articles, with colors representing topics (\cpt{STEM} and \cpt{Media}).}
\label{fig:2d_projections}
\end{figure*}

\section{RQ2: Shapes of Rhythms}
\label{sec:RQ2}

Answering RQ1, we established that the consumption pattern of articles exhibits daily regularities even after removing the weekly baseline rhythm. Next, we investigate the shape of these patterns.
Since the daily rhythm (wavelength 24 hours) was found to be the strongest by far, we henceforth focus on the daily instead of the weekly cycle; \ie, we now have $h \in \{0,\dots,23\}$ instead of $\{0,\dots,167\}$, with averages taken over 28 days instead of over four weeks.

\Figref{fig:STEMvsMovie_example} shows examples of the resulting daily time series for two articles $a$ associated with different topics (\cpt{STEM} and \cpt{Media}). \Figref{fig:STEMvsMovie_series} shows the shape of the normalized daily access volume time series $\Pr(h|a)$, and \Figref{fig:STEMvsMovie_ratio} shows their divergence $D_a(h)$ from the daily baseline rhythm $\Pr(h)$.

\subsection{Principal Components} 
To investigate prototypical article consumption behaviors, we extract the principal components describing the daily divergence time series $D_a(h)$ of article $a$. 
For this purpose, we stack the individual divergence time series $D_a(h)$ in a $439\text{K} \times 24$ matrix $D$ whose rows correspond to $a$ and whose columns correspond to $h$.
We mean-center $D$ column-wise and compute its singular value decomposition (SVD) $D=U \Sigma V^T$.
The (orthogonal) columns of $V$, then, are the principal components of $D$, capturing prototypical divergence time series.
Individual articles' divergence time series $D_a(h)$ can be approximated as linear combinations of the top principal components.
$\Sigma$ is a diagonal matrix whose $i$-th entry represents the standard deviation of the data points (rows of $D$) when projected on the $i$-th principal component.
Using the elbow method, we find that the first four principal components provide a good representation of consumption behavior, accounting, respectively, for 39.2\%, 19.7\%, 10.2\%, and 4.5\% of the total variance, jointly capturing 73.6\% of the total variance.

\Figref{fig:principal_components} shows the first four principal components of $D$.
Since each component can contribute positively or negatively to the reconstruction of the rows of $D$ (since principal components are sign-invariant), we depict each component in its positive and negative variant (dark and light).
The first and strongest component (PC1) captures the distinction between articles that receive more attention during the day versus evening/night (or vice versa), with two turning points around 7:00 and 21:00. The second component (PC2) describes patterns with a strong positive or negative contribution in the morning, with a peak around 5:00, which is reversed in the late hours of the day. The third principal component (PC3) captures a consumption pattern that peaks positively or negatively in the early morning and evening. Finally, the fourth principal component (PC4) models a similar behavior with the strongest contribution around 4:00 and in the early evening.

\subsection{Clustering}
Next, we aim to identify groups of articles with similar temporal access patterns via clustering.
As clustering tends to work better in low dimensions, we start by obtaining low-dimensional time series representations by truncating $U$ (obtained via SVD) to the first four columns $U_{1:4}$ (corresponding to projections on the first four principal components) and approximating article $a$'s divergence time series $D_a(h)$ via the $a$-th row of $U_{1:4}$.
We then cluster the rows of $U_{1:4}$ using $k$-means and search for the optimal number $k$ of clusters using the average silhouette width and sum-of-squares criteria. Both criteria indicate $k=2$ as the optimal number of clusters, suggesting that the articles cannot easily be separated into distinct groups. This intuition is supported by \Figref{fig:clusters}, which shows the clusters obtained for $k=2, 3, 4$. The upper plots show a UMAP \cite{mcinnes2018umap} reduction in two dimensions of the four principal components used to cluster the articles. The visualization is based on a sample of 10K random articles, color represents cluster assignments, and centroids are marked as black dots. 
The lower plots show the centroid of each cluster, reconstructed from the first four principal components.
These plots support the intuition that consumption behaviors do not separate into different groups but distribute along a continuum.%
\footnote{Silhouette scores obtained by grid-searching the number of principal components and clusters do not improve the separation. Density-based clustering yields the same conclusions.}

On the right side of each of the three scatter plots, we see articles popular during daytime. In contrast, on the left, we see articles popular during the evening\slash night. This partition becomes more evident when increasing the number of clusters. With three clusters, the data on the left separate into an additional group (green) containing articles popular during the night. Similarly, adding the fourth cluster isolates a set of articles (red) popular in the early morning.

\hfill

To summarize, investigating RQ2, we found that Wikipedia articles' access rhythms are driven by only few prototypical basis patterns (\eg, day \vs night).
Individual articles' access patterns can be captured as weighted combinations of those basis patterns, yet they do not cluster into distinct, well\hyp separated groups, but vary along a smooth continuum.

\section{RQ3: Correlates of Rhythms}
\label{sec:RQ3}
In this section, we investigate what factors correlate with the shape of daily consumption. We focus on three factors: article topics, access method, and user country.

\Figref{fig:STEMvsMedia} offers initial evidence that article topics are associated with daily access patterns. The plot represents the subset of articles from \Figref{fig:clusters} with topic \cpt{Media} or \cpt{STEM}. By coloring the data by topic (red for \cpt{Media}, blue for \cpt{STEM}), we can notice a natural separation akin to the clustering with $k=2$ clusters (\Figref{fig:clusters}, left). Note that the topic was not considered during clustering; the separation happens only based on the shape of access patterns. Articles about \cpt{STEM} are on the right side, indicating a pattern aligned with more attention during daytime, whereas articles about \cpt{Media} cover the right side, associated with evening and night consumption.

Going deeper, we
analyze prototypical shapes of access volume time series via regression analysis, quantifying the relationship of each factor (topics, access method, country) with time, and measuring the influence of each factor in an ablation study.

\begin{figure}[t]
    \centering
    \includegraphics[width=0.99\linewidth]{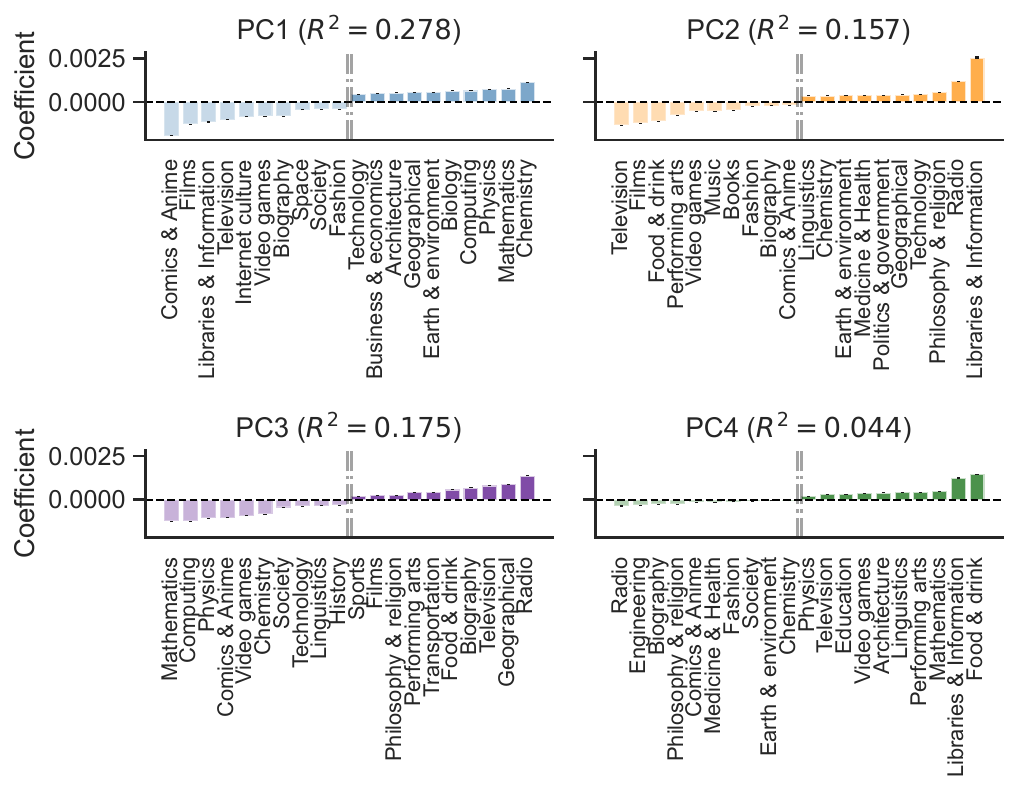}
    \caption{Linear regression coefficients of the topics most associated with each of the four main principal components of article access volume time series.}
    \label{fig:principal_components_topics}
    \vspace{-3mm}
\end{figure}

\begin{figure*}[t]
    \centering
    \includegraphics[width=0.99\linewidth]{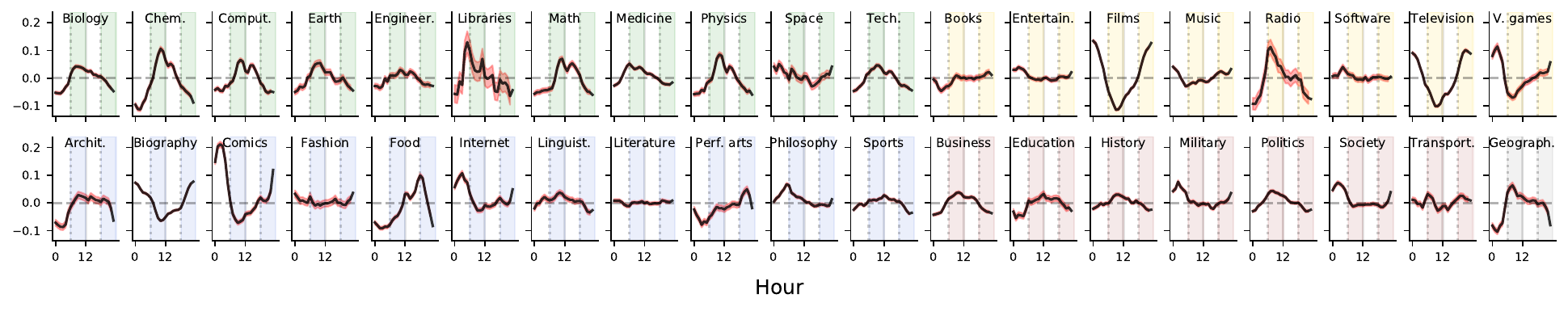}
    \caption{Linear regression coefficients of the interaction between topics and hour. Background colors represent top-level topics: \cpt{STEM} (green), \cpt{Media} (yellow), \cpt{Culture} (blue), \cpt{History \& Society} (red) and \cpt{Geographical} (gray). For each top-level topic, specific topics are sorted alphabetically. Red bands represent 95\% confidence intervals.}
    \label{fig:topic_coefficients_large}
\end{figure*}

\begin{figure*}[t]
    \centering
    \includegraphics[width=0.99\linewidth]{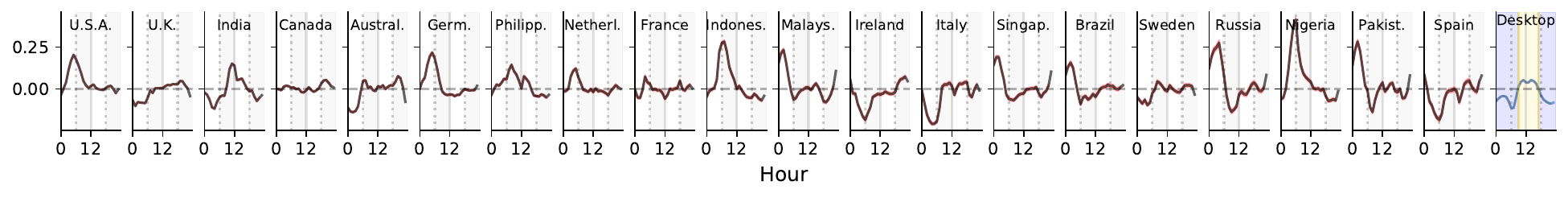}
    \caption{Linear regression coefficients of the interaction between country and hour (left, gray background, sorted by the total access volume), and between device and hour (rightmost plot).
    Yellow area in rightmost plot represents typical working hours in Western countries. Red bands represent 95\% confidence intervals. }
    \label{fig:country_desktop}
    \vspace{-3mm}
\end{figure*}

\subsection{Principal Components and Topics}

In an initial analysis, we leverage the principal components introduced earlier for RQ2
and investigate the relationship between topics and access patterns
using four linear regressions, one per top principal component. Given an article $a$, we predict the projection $U_{ai}$ of its divergence time series $D_a(h)$ onto the $i$-th principal component using $a$'s topics as binary predictors.

\Figref{fig:principal_components_topics} shows a summary of the coefficients obtained by fitting the four regressions.
The first principal component, capturing the day--night rhythm, separates articles about STEM from those about entertainment (also \cf\ \Figref{fig:STEMvsMedia}). 
Articles associated with topics such as \cpt{Comics \& Anime}, \cpt{Films}, and \cpt{Television} show a negative association with this principal component, suggesting nighttime popularity, whereas the temporal patterns of articles about \cpt{Chemistry}, \cpt{Mathematics}, and \cpt{Physics} show a positive association, suggesting daytime popularity.
The second component, associated with higher activity in the early morning and a drop during the evening, shows that the topics associated most positively with this pattern are \cpt{Libraries \& Information}, \cpt{Radio}, and \cpt{Philosophy \& religion}. On the other side of the spectrum, \cpt{Films}, \cpt{Television}, and \cpt{Food \& Drinks} are most negatively associated with this pattern (\ie, low activity in the early morning, spike during the evening).
Finally, the third and fourth principal components, showing peaks in the early morning and during the evening, are most positively associated with \cpt{Radio} and \cpt{Food \& drink}, respectively.

\subsection{Temporal Variation of Correlates}
\label{temporal_shapes}
Next, we are interested in the prototypical access volume time series associated with each level of each factor (topics, access method, user country).
A naive way of accomplishing this would be to first restrict the data to the respective factor level (\eg, requests for \cpt{Biology} articles only, or requests from Canada only) and then compute access volume time series for the restricted data only.
This approach is, however, compromised by the fact that the various factors contribute differently to the observed access pattern.
Thus, in order to disentangle the factors, we employ linear regression.


\xhdr{Model} We prepare the data as follows. First, we decompose each article $a$'s divergence time series $D_a(h)$ by country and access method. To avoid data sparsity issues, we limit ourselves to the 20 countries with the most pageloads. This step generates, for each article, 40 daily time series (20 countries, two access methods)
describing the
divergence of the attention to $a$ from a specific country and access method \vs\ the global baseline rhythm $\Pr(h)$. We explode each time series into 24 samples, 
such that
each article is represented by 960 data points describing its divergence from the baseline rhythm for each combination of 20 countries, two access methods, and 24 hours.
We then model divergence (transformed via the natural logarithm, to make the model multiplicative) as a linear combination of country, access method, topics, and hour of the day.
Additionally, since we are interested in modeling the relationship between country, access method, and topics \textit{with time}, we include interaction terms between each of these three factors with the hour-of-the-day indicator.
We represent all binary predictors via deviation coding, which lets us interpret the coefficients associated with each level as differences from the grand mean, which is itself captured by the intercept (since differences are taken in log space, they correspond to ratios with the grand mean in linear space).
With this setup, we keep the time series composed of at least 100 pageloads during the four-week period of study and fit a linear regression using a random sample of 30K articles.
The obtained model fits the data with $R^2=0.181$.

\xhdr{Coefficient analysis}
Sorting (by hour) the 24 interaction coefficients of each factor level with hour of the day allows us to characterize the typical temporal shape of each factor level (in terms of its deviation from Wikipedia's baseline rhythm) \textit{when controlling for all other factors}.

\Figref{fig:topic_coefficients_large} shows the temporal shapes of the topics organized in the five top-level topical groups. The plot shows how articles about \cpt{STEM} topics, such as \cpt{Chemistry}, \cpt{Physics}, and \cpt{Mathematics}, tend to receive more attention than average during the daytime and a visible reduction outside the typical working hours. On the other hand, articles about \cpt{Films}, \cpt{Television}, and \cpt{Biography} have an inverted shape, with less consumption than average during the day and a substantial increase during the evening. Interestingly, the shapes of the temporal patterns suggest that content about \cpt{Video Games}, \cpt{Comics \& Anime}, \cpt{Internet Culture}, \cpt{Military}, and \cpt{Society} are consumed by night-active readers, with relative consumption peaking during the night. 

On the contrary, articles about \cpt{Radio}, \cpt{Libraries \& Information}, and \cpt{Philosophy}, are more preferred in the early hours of the day.
Some of the shapes, especially the ones associated with \cpt{STEM} articles, show a reduction of attention around noon, suggesting that they might be affected by the lunch break when people's attention moves to other content types. This is corroborated by the fact that attention on articles about \cpt{Food} sees an increase during common meal times.

Next, \Figref{fig:country_desktop} shows the interaction coefficients of country by time. Some countries share similar daily patterns. 
For example, readers from the United States, Germany, Netherlands, and Nigeria tend to consume Wikipedia more than average in the early morning. This behavior is inverted for readers from India, Ireland, Italy, and Spain, who during the same hours consume less content than average. Meanwhile, other countries, such as Malaysia, Singapore, Brazil, Russia, and Pakistan, show higher consumption during the night. Furthermore, some countries, like the Philippines, Italy, France, and Spain, also reveal shared habits, such as a reduction of information consumption around noon, possibly associated with lunchtime.

Finally, the rightmost plot of \Figref{fig:country_desktop} shows the coefficients of the interaction of access method with time, in particular, the shape of the consumption from desktop devices. As already observed by breaking down the baseline rhythm by device in \Figref{fig:weekly_pattern_landing}, access from desktop devices is above the global average in the central hours of the day, between 9:00 and 17:00.

\begin{figure*}[t]

   \hfill
    \begin{minipage}[t]{.24\textwidth}
        \centering
        \includegraphics[height=2.94cm]{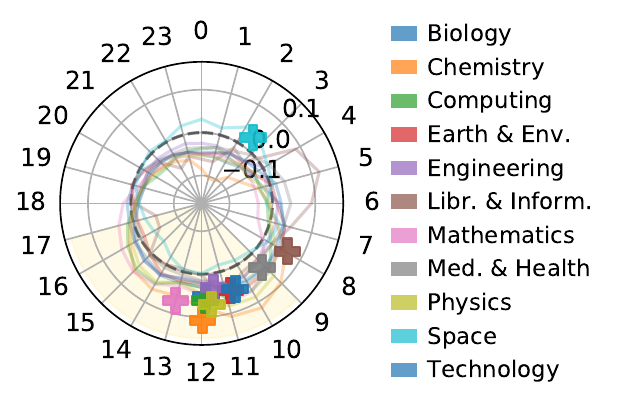}
        \subcaption{\cpt{STEM}}
        \label{fig:stem}
    \end{minipage}
       \hfill
    \begin{minipage}[t]{.24\textwidth}
        \centering
        \includegraphics[height=2.94cm]{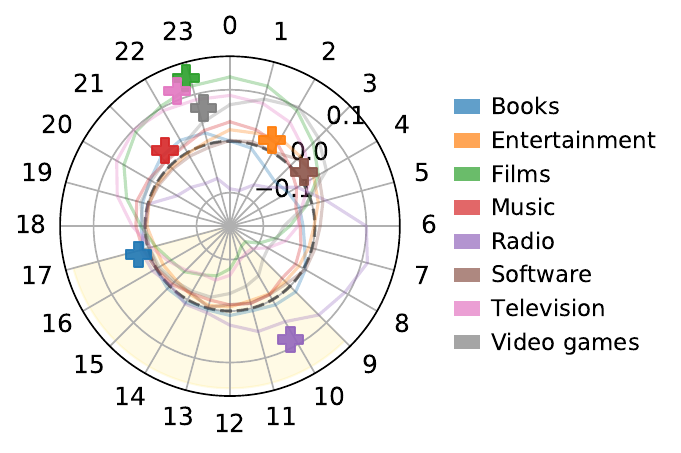}
        \subcaption{\cpt{Media}}
        \label{fig:media}
    \end{minipage}
       \hfill
    \begin{minipage}[t]{.24\textwidth}
        \centering
        \includegraphics[height=2.94cm]{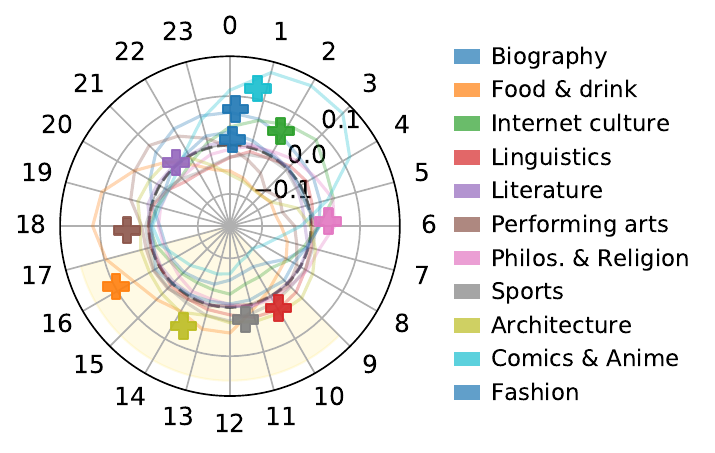}
        \subcaption{\cpt{Culture}}
        \label{fig:culture}
    \end{minipage}
       \hfill
    \begin{minipage}[t]{.24\textwidth}
        \centering
        \includegraphics[height=2.94cm]{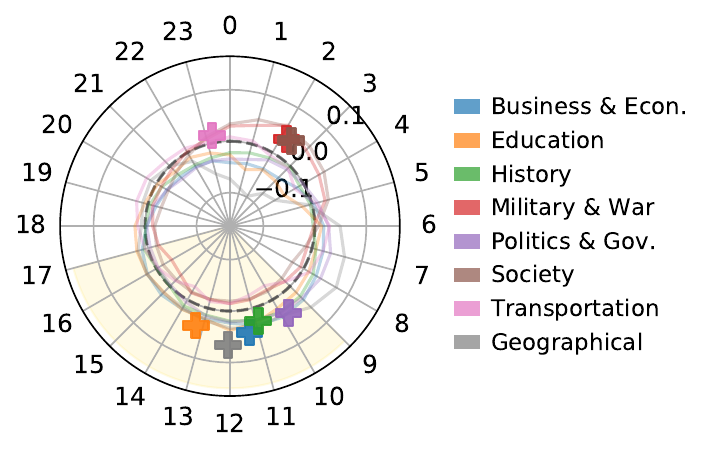}
        \subcaption{\cpt{History \& Society}}
        \label{fig:society}
    \end{minipage}

    \hfill
    
\caption{Typical access times of topics. Lines represent the coefficient time series of \Figref{fig:topic_coefficients_large}, and bold crosses the average times per topic. The dotted line marks the global baseline rhythm $\Pr(h)$. Yellow area shows typical working hours in Western countries. \cpt{Geographical} is included in \cpt{History \& Society} to compress the visualization.}
\vspace{-4mm}
\label{fig:average_time}
\end{figure*}

\subsection{Typical Access Times of Topics}
We now aim to summarize each topic-wise time series of \Figref{fig:topic_coefficients_large} concisely in a single number describing at what time of day articles from the respective topic are particularly popular.
In order to obtain point estimates of ``average time'', the simple arithmetic mean is unsuitable, given the circular nature of time (\eg, the average between 23:00 and 1:00 should be 0:00, not the arithmetic mean of 12:00),
so we use the angular mean instead (essentially an arithmetic mean in the 2D plane described by a clock's hands).
\Figref{fig:average_time} shows the topic-wise time series of \Figref{fig:topic_coefficients_large} in a circular fashion as light-colored curves, and average times as bold crosses, with one panel per top-level topic
(where a cross's closeness to the origin captures the variance of the respective time series).
We observe that articles about \cpt{STEM} (except for \cpt{Space}) tend to garner particularly much attention during working hours between 8:00 and 14:00. On the other hand, \cpt{Media} is, on average, consumed more during the evening and night. \cpt{Radio} is an exception in this group, peaking in the early morning, as visible in \Figref{fig:topic_coefficients_large}, bringing its daily average to around 10:00. Articles about \cpt{Books} are at the limit of the typical working hours, with average access in the late afternoon around 17:00.
Differently from the previous two groups, articles about \cpt{Culture} tend to have high variance and have average times spread over the day. Even in this case, entertainment topics such as \cpt{Fashion}, \cpt{Internet culture}, and \cpt{Comics \& Anime} are concentrated during the night, with average access after midnight.
Finally, content about \cpt{History \& Society} shows an average consumption split into two groups, night \vs\ working hours, with \cpt{Society}, \cpt{Military}, and \cpt{Transportation} concentrated during the late evening and night, and \cpt{Business \& economics}, \cpt{History}, and \cpt{Education} during the day.

%

\subsection{Strength of Correlates: Ablation Study}
Finally, we investigate each factor's influence on temporal access rhythms by estimating the strength of the interaction between each factor and time in an ablation study. Using the fitted model described above, we proceed as follows. We use a held-out dataset of 10K random articles not used in model\hyp fitting to assess the $R^2$ fit after permuting all the interaction terms of the factor being investigated. This approach, typically used to estimate feature importance in machine learning \cite{breiman2001random,fisher2019all}, has the advantage of keeping the model fixed---thus, no $R^2$ adjustment is required---and measuring the impact of removing the correlation between the selected feature and the dependent variable.

The original model has an $R^2$ fit of 0.181. When permuting the interaction terms of ``time by topic'', the $R^2$ drops to 0.055 (reduction by 69\%);
when permuting ``time by access method'', the $R^2$ is 0.040 (reduction by 78\%); and finally, when permuting ``time by country'', the $R^2$ score is $-0.0298$%
\footnote{A negative $R^2$  is possible on a held-out set, indicating that a flat line approximates the data better than the model.} 
(reduction of 116\%). A permutation of all three factors reduces the $R^2$ to $-0.429$. These observations suggest that all factors are essential for the prediction and indicate that the interaction of time with the readers' country plays the most important role.

\hfill

To summarize, in answering RQ3, we found that
temporal regularities in the popularity of Wikipedia articles vary in important and systematic ways
by topic, access method, and user country,
and that, out of these, user country has the strongest influence.

\section{Discussion}
\label{sec:discussion}
\xhdr{Summary of findings}
In this study, we conducted the first large-scale study on temporal rhythms of information consumption on Wikipedia, based on a new timezone\hyp corrected dataset of hourly-aggregated pageviews where the reader's local time was inferred from Wikipedia's webrequest logs.

First, we showed that the overall information consumption exhibits daily rhythmic patterns, with the strongest components having periods of 24 and 12 hours, respectively. We further showed that the consumption of individual articles follows specific rhythms that cannot be explained by Wikipedia's overall wake--sleep baseline rhythm. Rather, each article reveals a specific consumption fingerprint throughout the day.

Second, we provided a systematic description of the principal shapes of the different rhythms of the individual articles. We find that the main shape distinguishes articles that are read disproportionally more during the day than during the night (and vice versa). We do not, however, find distinct clusters of shapes but, instead, observe a continuum of different rhythms of information consumption. 

Third, we show systematic differences in consumption patterns based on the reader's context (country or device) or the article's topics. We showed that articles of specific topics are associated with specific rhythmic patterns. For example, articles on \cpt{STEM} and \cpt{Media} naturally separate into two clusters related to rhythms with disproportional attention during the day and night, respectively. More generally, this leads to markedly different ``average'' times when articles from different topics are accessed. We also showed that pageloads from mobile \vs\ desktop devices are driven by substantially different 24-hour rhythms, with mobile pageloads showing an almost twofold increase in the evening hours compared to the day, which is absent for desktop devices. Lastly, we also find substantial variation in the access patterns across different countries.


\xhdr{Implications}
Our work shows that context is an important element to consider when trying to understand how information on Wikipedia is consumed. This has several implications.

\textit{Diversity of information needs.} Wikipedia as a platform fulfills multiple information needs. 
These needs vary not only with geographic location, such as country~\cite{lemmerich_why_2019}, but also with time of day when a reader accesses the platform. In order to serve these needs, we need to consider the heterogeneity of Wikipedia's audience in \textit{space and time}. Additionally, given the extension of topics that Wikipedia covers and its large adoption, our study offers insights into what content people consume online during the day, at a scale generally accessible only to search engines and Internet providers, with implications for design beyond Wikipedia.

\textit{Cultural diversity.}
Given the diversity of Wikipedia access rhythms across countries  (\Figref{fig:country_desktop}),
future work could revisit our results from an anthropological perspective, \eg, in order to pinpoint cultural differences in the daily rhythm of information needs---or curiosity---across the globe.
For instance, we are fascinated by the fact that the rhythm of (English) Wikipedia's popularity in the U.S.\ is nearly exactly opposed to that in Spain, and that India and Pakistan---two countries that once were one---have entirely different rhythms of English Wikipedia consumption.
Wikipedia logs offer a window into where in the world people care about what, when, and we anticipate that studying access time series while accounting for the interaction of country by topic (which we omitted here) has enormous potential for understanding global patterns of needs for knowledge.

 \textit{Metrics for information needs.} The popularity of an article (\ie, pageloads) is often used as an indicator for its relevance~\cite{wulczyn_growing_2016} or a covariate used for stratification in the analysis of articles \cite{piccardi2020quantifying}. In our context, this corresponds to only looking at the volume of the consumption pattern and ignoring its shape throughout the day. In contrast, our approach focuses on the shape of the consumption pattern revealing substantial differences between articles. Thus, using the total number of pageviews (volume) as a single relevance metric will likely miss nuances about how these articles are useful to the reader. Complementary metrics capturing the usage of articles, such as the shape, describe important properties that might act as confounding factors, and other studies should consider them in stratified analyses. 
In addition, they could also be directly useful for editors providing additional information about the audience of articles, potentially helping to bridge the gap reported in the misalignment between supply and demand of articles \cite{warncke2015misalignment}.

\textit{Customization.}
 These results should be taken into account when building tools to support users in finding and accessing relevant information (\eg Wikipedia's search or recommendations such as RelatedArticles\footnote{\url{https://www.mediawiki.org/wiki/Extension:RelatedArticles}}) or to customize their online experience
 . For example, on average offering information about a movie in the morning has less value than in the evening. Similarly, understanding the content that draws more attention at a different time of the day has implications for designing systems to engage potential editors when they are more inclined to contribute to specific content.

\textit{Infrastructure optimization.} As observed in previous work \cite{reinoso2012characterization}, these findings are valuable for optimizing the Wikimedia infrastructure. Given the scale of Wikipedia, optimizing data caching and load-balancing to reflect the actual needs of the consumers across space and time can offer significant benefits for the platform's performance.

\xhdr{Limitations and future work} 
Our analysis may have some data limitations, although we argue that it is unlikely they would alter our main findings and conclusions. For example, despite the best efforts of the Wikimedia infrastructure team in developing heuristics to detect bots, the data may still contain some automated traffic. Similarly, country and timezone are identified using the IP address of the request, which may be sensitive to the use of VPNs. Other aspects that may influence our findings are unobserved external factors such as biases introduced by search engines. Since the large majority of Wikipedia traffic originates from search engines \cite{piccardi2023large}, search engine results varying with the time of the day would also impact visits to Wikipedia.

One concrete limitation of our study that we aim to address in future work is focusing only on the English edition of Wikipedia. Readers from non-English-speaking countries consuming content from this edition may be a biased population. Future work should extend our study to multiple languages to compare information behaviors worldwide.
Also, future work should study how individual\hyp level (as opposed to population\hyp level) information needs change during the day, possibly combined with demographic data, such as age or education \cite{kulshrestha2020web}. In our case study, this was not possible due to privacy constraints.

Another limitation of this study is the potential impact of external factors. The readers' behavior could be affected by extended temporal cycles, including annual trends or worldwide occurrences, such as the mobility restrictions imposed during the COVID-19 pandemic \cite{ribeiro2020sudden}. Consequently, further research should explore these patterns over a period extending beyond the four-week duration of this study.

Moreover, future work should explore the shape and strength of temporal rhythms on other platforms beyond Wikipedia, especially on search engines, which tend to be the first resource we turn to when we need information online. In this sense, studying search engines can offer a complementary view to our findings.
Ultimately, we hope that our study can help pave the way to better serve the needs of Wikipedia readers and Web users in general. 

\xhdr{Ethical considerations} Server logs may contain sensitive information with implications for the privacy of users. In this work, we pay special care to ensure the researchers access only anonymized records, excluding activities of logged-in users and editors that may be linked through public data. Our findings describe aggregated behaviors that represent a minimal risk for privacy violations. We believe the benefit of presenting them outweighs the potential risks.

\section{Acknowledgements} We would like to thank Leila Zia and Brandon Black for insightful discussions and for reviewing an initial draft of this paper. West's lab is partly supported by grants from
Swiss National Science Foundation (200021\_185043, TMSGI2\_211379),
Swiss Data Science Center (P22\_08),
H2020 (952215),
Microsoft Swiss Joint Research Center,
and Google,
and by generous gifts from Facebook, Google, and Microsoft. Tiziano Piccardi is supported by the Swiss National Science Foundation (Grant P500PT-206953).

\bibliography{References}

\subsection{Ethics Checklist}

\begin{enumerate}

\item For most authors...
\begin{enumerate}
    \item  Would answering this research question advance science without violating social contracts, such as violating privacy norms, perpetuating unfair profiling, exacerbating the socio-economic divide, or implying disrespect to societies or cultures?
    \answerYes{Yes.}
  \item Do your main claims in the abstract and introduction accurately reflect the paper's contributions and scope?
    \answerYes{Yes.}
   \item Do you clarify how the proposed methodological approach is appropriate for the claims made? 
    \answerYes{Yes. The sections describing the research questions introduce the methods and the relative explanations.}
   \item Do you clarify what are possible artifacts in the data used, given population-specific distributions?
    \answerYes{Yes. See Data.}
  \item Did you describe the limitations of your work?
    \answerYes{Yes. See Discussion.}
  \item Did you discuss any potential negative societal impacts of your work?
    \answerYes{Yes. See Ethical considerations. We do not foresee any significant negative impact.}
      \item Did you discuss any potential misuse of your work?
    \answerYes{Yes. See Ethical considerations. We do not foresee any significant negative impact.}
    \item Did you describe steps taken to prevent or mitigate potential negative outcomes of the research, such as data and model documentation, data anonymization, responsible release, access control, and the reproducibility of findings?
    \answerYes{Yes. See Data.}
  \item Have you read the ethics review guidelines and ensured that your paper conforms to them?
    \answerYes{Yes.}
\end{enumerate}

\item Additionally, if your study involves hypotheses testing...
\begin{enumerate}
  \item Did you clearly state the assumptions underlying all theoretical results?
    \answerNA{N/A}
  \item Have you provided justifications for all theoretical results?
    \answerNA{N/A}
  \item Did you discuss competing hypotheses or theories that might challenge or complement your theoretical results?
    \answerNA{N/A}
  \item Have you considered alternative mechanisms or explanations that might account for the same outcomes observed in your study?
    \answerNA{N/A}
  \item Did you address potential biases or limitations in your theoretical framework?
    \answerNA{N/A}
  \item Have you related your theoretical results to the existing literature in social science?
   \answerNA{N/A}
  \item Did you discuss the implications of your theoretical results for policy, practice, or further research in the social science domain?
    \answerNA{N/A}
\end{enumerate}

\item Additionally, if you are including theoretical proofs...
\begin{enumerate}
  \item Did you state the full set of assumptions of all theoretical results?
    \answerNA{N/A}
	\item Did you include complete proofs of all theoretical results?
    \answerNA{N/A}
\end{enumerate}

\item Additionally, if you ran machine learning experiments...
\begin{enumerate}
  \item Did you include the code, data, and instructions needed to reproduce the main experimental results (either in the supplemental material or as a URL)?
    \answerNA{N/A}
  \item Did you specify all the training details (e.g., data splits, hyperparameters, how they were chosen)?
    \answerNA{N/A}
     \item Did you report error bars (e.g., with respect to the random seed after running experiments multiple times)?
    \answerNA{N/A}
	\item Did you include the total amount of compute and the type of resources used (e.g., type of GPUs, internal cluster, or cloud provider)?
    \answerNA{N/A}
     \item Do you justify how the proposed evaluation is sufficient and appropriate to the claims made? 
    \answerNA{N/A}
     \item Do you discuss what is ``the cost`` of misclassification and fault (in)tolerance?
    \answerNA{N/A}
  
\end{enumerate}

\item Additionally, if you are using existing assets (e.g., code, data, models) or curating/releasing new assets...
\begin{enumerate}
  \item If your work uses existing assets, did you cite the creators?
    \answerNA{N/A}
  \item Did you mention the license of the assets?
    \answerNA{N/A}
  \item Did you include any new assets in the supplemental material or as a URL?
    \answerNA{N/A}
  \item Did you discuss whether and how consent was obtained from people whose data you're using/curating?
    \answerNA{N/A.}
  \item Did you discuss whether the data you are using/curating contains personally identifiable information or offensive content?
    \answerNA{N/A.}
\item If you are curating or releasing new datasets, did you discuss how you intend to make your datasets FAIR?
\answerNA{N/A.}
\item If you are curating or releasing new datasets, did you create a Datasheet for the Dataset)? 
\answerNA{N/A}
\end{enumerate}

\item Additionally, if you used crowdsourcing or conducted research with human subjects...
\begin{enumerate}
  \item Did you include the full text of instructions given to participants and screenshots?
    \answerNA{N/A}
  \item Did you describe any potential participant risks, with mentions of Institutional Review Board (IRB) approvals?
    \answerNA{N/A}
  \item Did you include the estimated hourly wage paid to participants and the total amount spent on participant compensation?
    \answerNA{N/A}
   \item Did you discuss how data is stored, shared, and deidentified?
   \answerYes{Yes. See Data.}
\end{enumerate}

\end{enumerate}

\end{document}